\documentclass[ preprint]{aastex61}

\usepackage{epsfig,graphicx,amssymb,epstopdf,mathrsfs,subfigure, amsmath}

\def\arcdeg{\hbox{$^\circ$}}

\def\simlt{\mathrel{\hbox{\rlap{\hbox{\lower4pt\hbox{$\sim$}}}\hbox{$<$}}}}
\def\simgt{\mathrel{\hbox{\rlap{\hbox{\lower4pt\hbox{$\sim$}}}\hbox{$>$}}}}

\newcommand{\vFv}{\mbox{$\nu F_{\nu}$}}

\submitjournal{ApJL}
\shorttitle{GBM Bayesian sGRB Catalog}
\shortauthors{Burgess et al. }
\begin{document}

\title{A Bayesian Fermi-GBM Short GRB Spectral Catalog}

\author[0000-0003-3345-9515]{ J. Michael Burgess}
\affiliation{Max-Planck-Institut f{\"u}r extraterrestrische Physik,
  Giessenbachstrasse, D-85748 Garching, Germany}
\affiliation{Excellence Cluster Universe, Technische Universit\"{a}t
      M\"{u}nchen,  Boltzmannstra{\ss}e 2, 85748, Garching, Germany}
\author[0000-0003-1256-173X]{J. Greiner}
\affiliation{Max-Planck-Institut f{\"u}r extraterrestrische Physik,
  Giessenbachstrasse, D-85748 Garching, Germany}

\author[0000-0003-4477-1846]{D. B\'egu\'e}
\affiliation{Max-Planck-Institut f{\"u}r extraterrestrische Physik,
  Giessenbachstrasse, D-85748 Garching, Germany}

\author[0000-0001-8747-0627]{F. Berlato}
\affiliation{Max-Planck-Institut f{\"u}r extraterrestrische Physik,
  Giessenbachstrasse, D-85748 Garching, Germany}
\affiliation{Physik Department, Technische Universit\"at M\"unchen, D-85748 Garching, 
James-Franck-Strasse, Germany}

\correspondingauthor{J. Michael Burgess}
\email{jburgess@mpe.mpg.de}

%%%%%%%%%%%%%%%%%%%%%%%%%%%%%%%%%%%%%%%%

\begin{abstract}
  With the confirmed detection of short gamma-ray burst (GRB) in
  association with a gravitational wave signal, we present the first
  fully Bayesian {\it Fermi}-GBM short GRB spectral catalog. Both peak
  flux and time-resolved spectral results are presented. Additionally,
  we release the full posterior distributions and reduced data from
  our sample. Following our previous study, we introduce three
  variability classes based of the observed light curve structure.
\end{abstract}

% Up to 6 keywords

%%%%%%%%%%%%%%%%%%%%%%%%%%%%%%%%%%%%%%%%

\section{Introduction}
\label{sec:intro}
The {\it Fermi} Gamma-ray Burst Monitor (GBM) \citep{Meegan:2009} is
the most prolific detector of short gamma-ray bursts (GRBs). Over its
nine year mission beginning on 2008, July, GBM has detected over 300
short GRBs. Long believed to be the byproduct of binary neutron star
mergers, the recent association of GW170817
\citep{Abbott:2017,Abbott:2017it} with the short GRB 170817A
\citep{Goldstein:2017hm} has made the study of GBM short GRB
population properties pertinent. First of all, the low luminosity
combined with otherwise typical spectral properties of short GRBs
demands an explanation of the physical emission mechanism
\citep[e.g.][see our accompanying
today]{Kasliwal:2017,Begue:2017}. Next, given the detection rate
within LIGO-O2 and consistent predictions from population studies
\citep[e.g.][]{Chruslinska:2017}, it is obvious to ask for the
detection of similar events in the GBM archival data which remained
unrecognized as nearby mergers \citep[e.g.][]{Burgess:2017aa}. This
also includes the question as to whether or not it is possible to
identify nearby binary neutron star mergers based on just the
gamma-ray data and potential optical/NIR follow-up?  Last but not
least, questions like a clear distinctive separation from
long-duration GRBs (based on the hardness-duration, lags, temporal
properties), the physical interpretation of soft tails or the
relations to magnetars \citep[e.g.][]{Rowlinson:2014} all require
input from a homogeneously deduced sample of spectral parameters.

Past GBM spectral catalogs
\citep[e.g.][]{Goldstein:2012aa,Gruber:2014aa,Yu:2016aa} have utilized
maximum-likelihood methods to provide spectral properties of GRBs to
the community. Herein, we have invoked Bayesian analysis to extract
both the temporal and spectral properties of short GRBs. This allows
for the injection of our prior beliefs about the properties of short
GRBs which results in the ability to uniformly model the data across
various photon functions. Additionally, we provide the results of our
analysis and data reduction to the community to encourage followup
studies.

This letter is organized as follows. First, we describe the sample
selection and data reduction. Next, we detail out spectral fitting
procedure and the catalog distributions. Finally, we briefly discuss
the implications of our results.

\section{Sample Selection}
\label{sec:sample}

The Fermi Science Support
Center\footnote{https://fermi.gsfc.nasa.gov/ssc/} (FSSC) provides
public data from the Fermi mission including GBM burst
data\footnote{https://heasarc.gsfc.nasa.gov/FTP/fermi/data/gbm/bursts/}. Additionally,
the GBM public GRB burst
catalog\footnote{https://heasarc.gsfc.nasa.gov/W3Browse/fermi/fermigbrst.html}
provides up-to-date durations and background selections for all
triggers classified as GRBs since the beginning of the mission.

Using these databases, we selected all GRBs with a T$_{90}$ duration
less than 7 seconds and retrieved the time-tagged event (TTE) data,
response matrices and background selections for detectors with a
viewing angle less than 60$^{\circ}$ from the reported source
location. While we will eventually use GRBs with a duration less than
2 seconds, we obtain those with a longer T$_{90}$ because we will use
a different duration measure as detailed in Section
\ref{sec:method}. GBM releases response matrices in two forms. Some
GRBs have responses for a single time interval (RSP) and others have
responses determined for multiple time intervals to account for the
slewing of the spacecraft (RSP2). If RSP2 files are available, we
utilize these over the RSP files.

The original sample includes 543 GRBs before data reduction. Due to
issues with background selections and lack of significant signal, some
GRBs were removed from the sample resulting in 321 short GRBs and 525
time-resolved spectra. The details of the data reduction are discussed
in the following section.

\section{Methodology}
\label{sec:method}
Following our work in \citet{Burgess:2017aa}, we apply a uniform
methodology for background fitting, temporal binning, and
time-resolved source selection. Each step is detailed in the following
paragraphs.

\subsection{Data Reduction}

For each GRB in our sample, an off-source background interval is
selected using the intervals identified in the GBM online
catalog. Using these intervals, a series of four polynomials of
increasing degree (from 0-3) is fit to the total rate in time. The
likelihood for the fit is unbinned Poisson and as nested models, a
simple likelihood ratio test (LRT) is applied to find the optimum
polynomial degree without over-fitting. With the degree determined, a
polynomial of the same degree are fit to each of the 128 PHA channels
to estimate the background model for the rate in that channel. The
background will be integrated in time over each source interval. As
the background estimation is the result of a maximum-likelihood fit,
the errors are assumed to be Gaussian distributed. Thus, when
calculating the statistical error on the background for each channel,
the covariance matrix of the background fit is propagated into the
temporal integration resulting in our background error,
$\sigma_{\rm b}$.

With the background fitted, we apply Bayesian blocks
\citep{Scargle:2013} to the temporally unbinned source interval for
each detector (T$_0 -5+10$ s) with a chance probability parameter of
$p_0=0.01$. The background model is used to shift the background from
a non-homogeneous Poisson process to a homogeneous one. If no change
points are inferred, the GRB is discarded from the sample. The
detector light curve with the highest rate significance over the
background is selected and its inferred change points are mapped to
all other detectors. We note that the appropriate significance measure
to use is via a likelihood ratio similar to that derived by
\citet{Li:1983}. However, that likelihood is derived for the
significance of one Poisson rate over another. Since our determined
background model possesses Gaussian errors, the appropriate likelihood
ratio is that where we seek an excess over a Gaussian background
rate. Thus, we determine significance via the method of
\citet{Vianello:2017}.

The intervals for spectral analysis are now selected by retaining all
bins with a significance greater than 3 $\sigma$. For each bin, the
background model is integrated over the bin's time bounds and a source
and background are exported to PHA files. Similarly, if the GRB has
an RSP2 file, a weighted response matrix is calculated and exported.

\subsection{Spectral Analysis}

For each temporal bin, we fit both a Band function
\citep{Band:1993,Greiner:1995}

\begin{equation}
  \label{eq:band}
   F \left(E \right)\;=\;K
  \begin{cases}
    {\left(\frac{E}{\rm 100\;keV}\right)^{\alpha} \exp\left(-\frac{E}{E_{\rm cut}}\right)} &
{E\le (\alpha-\beta) E_{\rm cut} }\\
    {\left(\frac{E}{\rm 100\;keV}\right)^{\beta}\exp {\left(\beta-\alpha\right)}\left[\frac{(\alpha-\beta)E_{\rm
cut}}{{\rm 100\;keV}}\right]^{\alpha-\beta}}
    & {E>(\alpha-\beta) E_{\rm cut}  }
  \end{cases}  
\end{equation}

\noindent
and a cutoff power law function
\begin{equation}
  \label{eq:cpl}
  F \left(E \right)\;=\;K \left(\frac{E}{\rm 100\;keV}\right)^{\alpha} \exp\left(-\frac{E}{E_{\rm cut}}\right)\text{.}
\end{equation}
\noindent
Both functions are parameterized in terms of a cutoff energy
($E_{\rm cut}$) rather than a $\vFv$-peak energy to reduce
correlations between the peak and the low-energy spectral index
($\alpha$). We do not fit a simple power law model to the data because
we expect a spectral peak somewhere within the GBM energy range as
higher spectral peaks have never been observed. When power law models
are fit to short GRBs, it is typically found that the photon spectral
index is $\sim>-2$ which would imply a peak outside the GBM spectral
window. As we discuss below, we mitigate this with our prior
choices. For all parameters except the photon model normalization, we
adopt informative priors from previous catalogs. In particular, we
choose a normal prior on the cutoff energy of both the Band and CPL
functions centered at 200 keV. Thus, unless the data are more
informative than the prior, we impose that a spectral peak exist in
the GBM spectral window. The following prior choices were used:
\begin{eqnarray}
\alpha &\sim& \mathcal{N}\left(\mu=-1. , \sigma = 0.5\right) \\
E_{\rm cut} &\sim& \mathcal{N}\left(\mu=200. , \sigma = 300\right) \\
\beta &\sim& \mathcal{N}\left(\mu=-2.25 , \sigma = 0.5\right)\text{.}
\end{eqnarray}

The spectra are fit via a Poisson-Gaussian likelihood to account for
Poisson distributed total counts and Gaussian distributed background
estimate. This profile-likelihood removes the need for a background
spectral model by essentially assuming a parameter in each spectral
bin and profiling it out. This requires at least one background count
in each spectral bin and thus, we bin the spectra to achieve this
goal. However, this rarely reduces the number of bins by more than one
or two.

To account for systematics in the GBM responses, we scale all
responses except one by a normalization constant, a so-called
effective area correction. A similar procedure was used in
\citet{Yu:2016aa}. The GBM responses are claimed accurate to with 10\%
\citep{Bissaldi:2009}, and therefore we place a Cauchy prior centered
at unity with a 10\% standard deviation on these normalization
constants. The Cauchy prior is informative in its tails but allows for
some lack of certainty around the mean.  These corrections will be
marginalized into our spectral parameter posteriors.

Finally, to perform the spectral fit, we sample the posterior with the
{\tt MULTINEST} algorithm \citep{Feroz:2009}. For each fit, 600 live
points are used. {\tt MULTINEST} ceases to sample when a tolerance on
the marginal likelihood integral has been achieved. Hence, we record
the value of the marginal likelihood ($Z$) for each fit. Due to our
use of informative priors, we can employ model selection between the
Band and CPL functions via marginal likelihood ratios. This equates
to choosing the Band function over CPL if the information in the data
for the high-energy spectral index ($\beta$) contains more information
than the prior. Model selection is performed per intra-burst interval
allowing for the spectral model to evolve within a burst.

All temporal and spectral analysis is carried out with the
Multi-Mission Maximum Likelihood framework
\citep[3ML;][]{Vianello:2015}. The results are stored in Flexible Image
Transport System (FITS) "analysis results" files which are readable by
3ML or any normal FITS reader. They contain information regarding the
spectral model and the full posterior of all parameters. Each file can
be used to fully setup the analysis for replication. Additionally, we
propagate the spectral fit into an energy flux (F$_{\rm E}$)
calculation integrated over the 10 keV - 1 MeV energy range resulting
in a marginal distribution for the energy flux. Note that via the
analysis result FITS files, the fluxes can be recomputed over any
energy range.

In the past we have argued that a more complicated spectral fitting
algorithm should be invoked to account for systematics in the
locations of GRBS \citep[see the BALROG][]{Burgess:2016aa} and that
physical photon models provide better insight into the emission
mechanism of GRBs \citep{Burgess:2015aa,Burgess:2017wp}. However, work
on the systematics in location are ongoing, and as we note in
\citet{Begue:2017}, the emission mechanism of short GRBs requires
further modeling. Therefore, with the confirmation of binary neutron
star mergers as at least one progenitor of short GRBs, we find it
pertinent to proceed with the classical photon models in order to
provide the modeling community with an empirical view of the spectra.

\section{Results}

\begin{figure}[h]
  \centering
  \figurenum{1}
  \includegraphics[width=0.45 \textwidth]{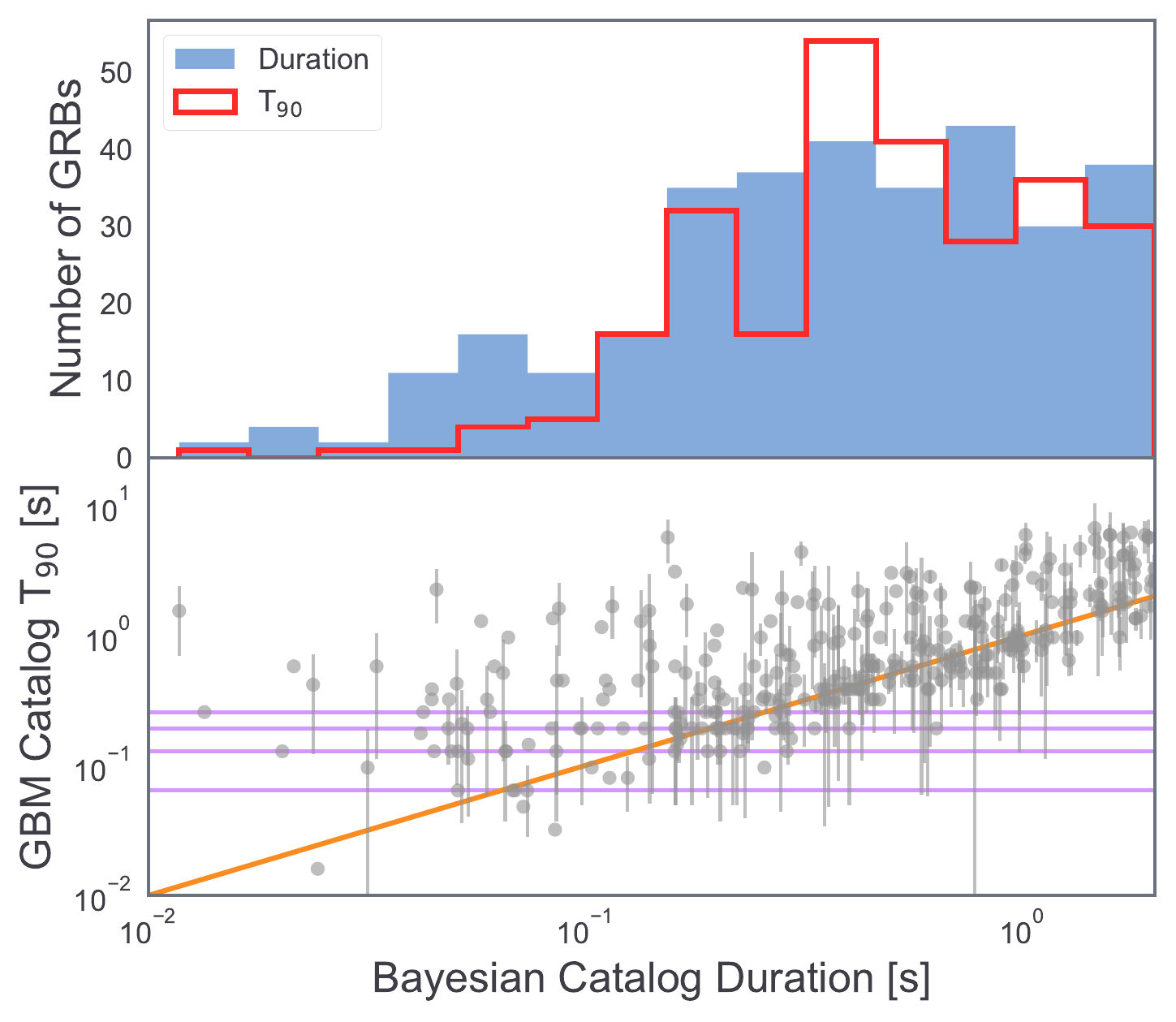}
  \caption{(top) Our duration and the GBM T$_{90}$
    distributions. (bottom) Comparison of the two duration measures
    including the errors from the GBM T$_{90}$ measurements. The
    purple lines indicate four multiples of the CTIME temporal binning
    of 0.064 s. The orange line demonstrates a one-to-one
    correspondence.}
  \label{fig:t90}
\end{figure}

From our temporal binning and spectral fitting results, we provide the
joint and individual parameter distributions for our sample. We
consider the total duration of the emission as the interval from the
beginning of the first to ending of the last Bayesian block from our
temporal analysis. We note that this is quite different than the
typical T$_{90}$ \citep{Koshut:1995} used for GRB durations and thus,
we simply call this the duration of the GRB. Our duration does not
account for the detector response as it is calculated in count space,
however, we do not compare our results to previous duration measures
for any interpretation. Moreover, duration measures are somewhat
arbitrarily performed in different energy ranges, and differ across
instruments. What is perhaps most important is how relative durations
vary within a sample. A comparison of T$_{90}$ and our duration is
shown in Figure \ref{fig:t90}. The deviation at short durations
between the two measures is the result of many of the GBM T$_{90}$
being quantized to multiples of the temporal binning (0.064 s) of the
CTIME data \citep{Meegan:2009}. Our measure is computed on the
unbinned TTE data and thus is limited only by the 2 $\mu$s clock of
the GBM DPU.

In \citet{Burgess:2017aa}, we classified light curves into three
classes: simple (consisting of only one significant bin), pulse-like
(consisting of several significant contiguous bins) and complex
(consisting of non-contiguous significant bins). We will examine the
parameter distributions of both the combined and individual classes
below.

The selection of the Band function over the CPL was made if $2 \ln K$
of the Band function was greater than 10 that of the CPL. Here,
$K = Z_{\rm Band}/ Z_{\rm CPL}$. With this
criterion, only 12 of the time-resolved spectra were best
fit by the Band function and 513 were best fit by the CPL function.

\subsection{Parameter Distributions}

\begin{figure*}[htb!]
\centering
\figurenum{2}  \subfigure{\includegraphics[scale=.4]{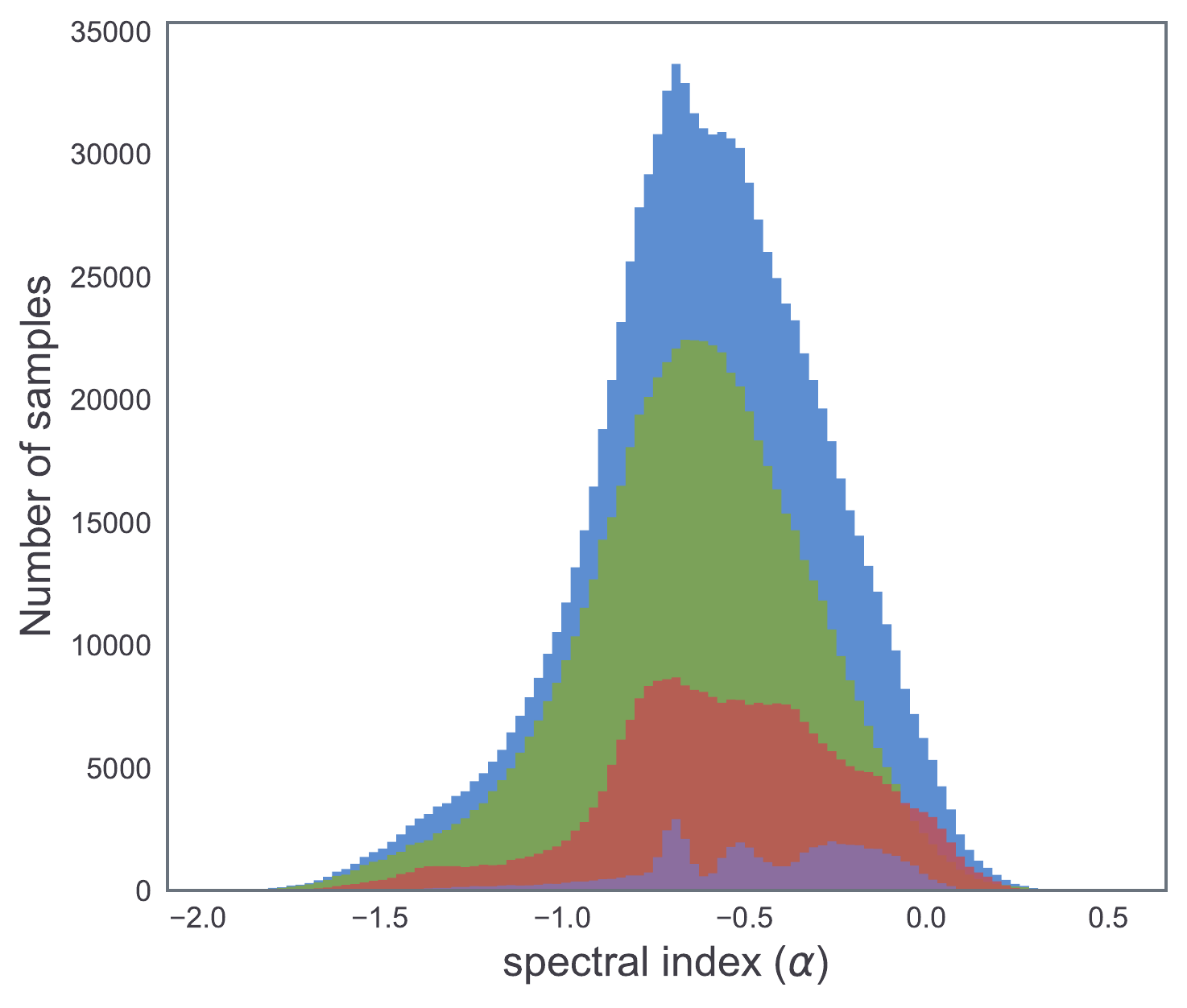}}\subfigure{\includegraphics[scale=.4]{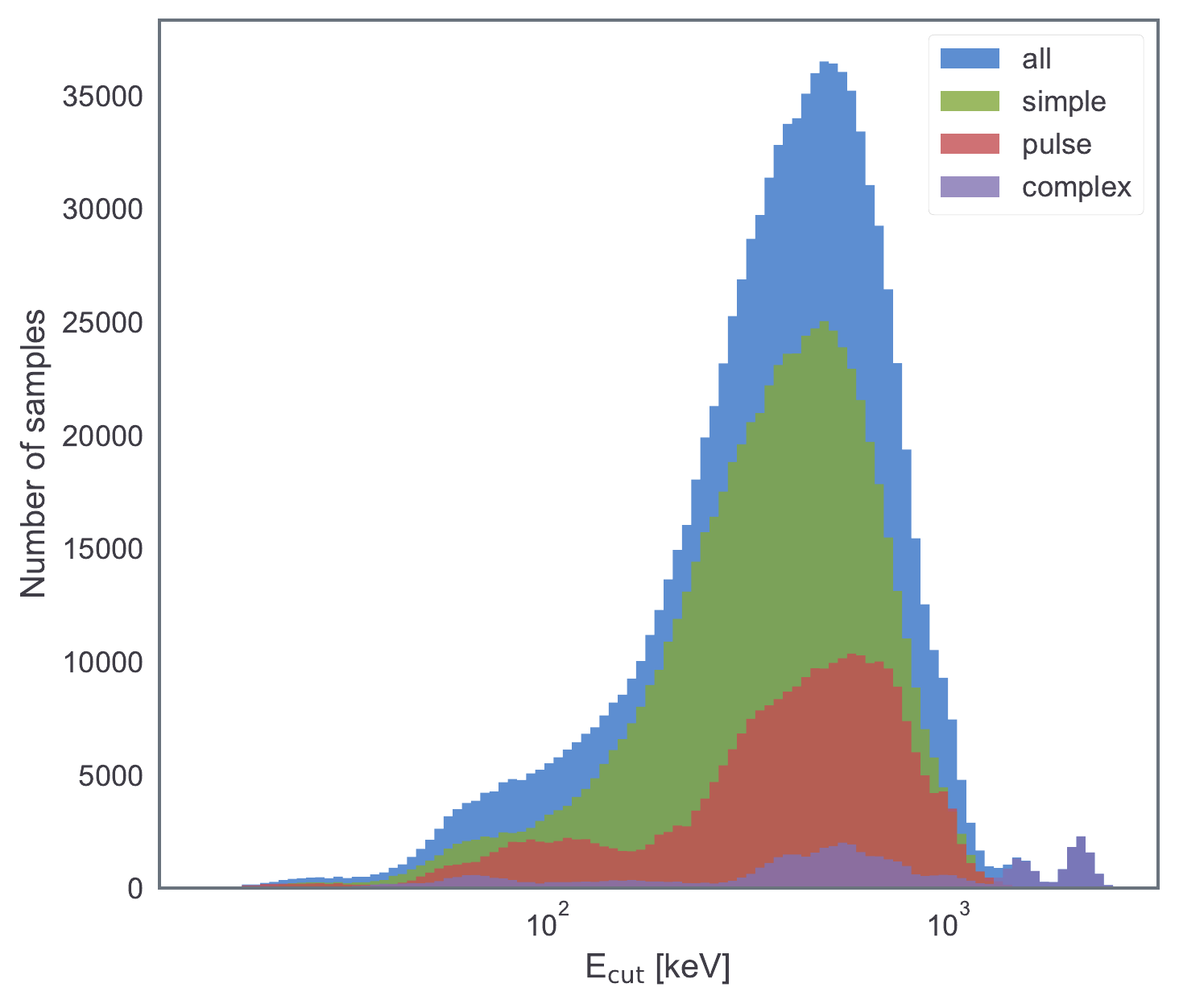}}\subfigure{\includegraphics[scale=.4]{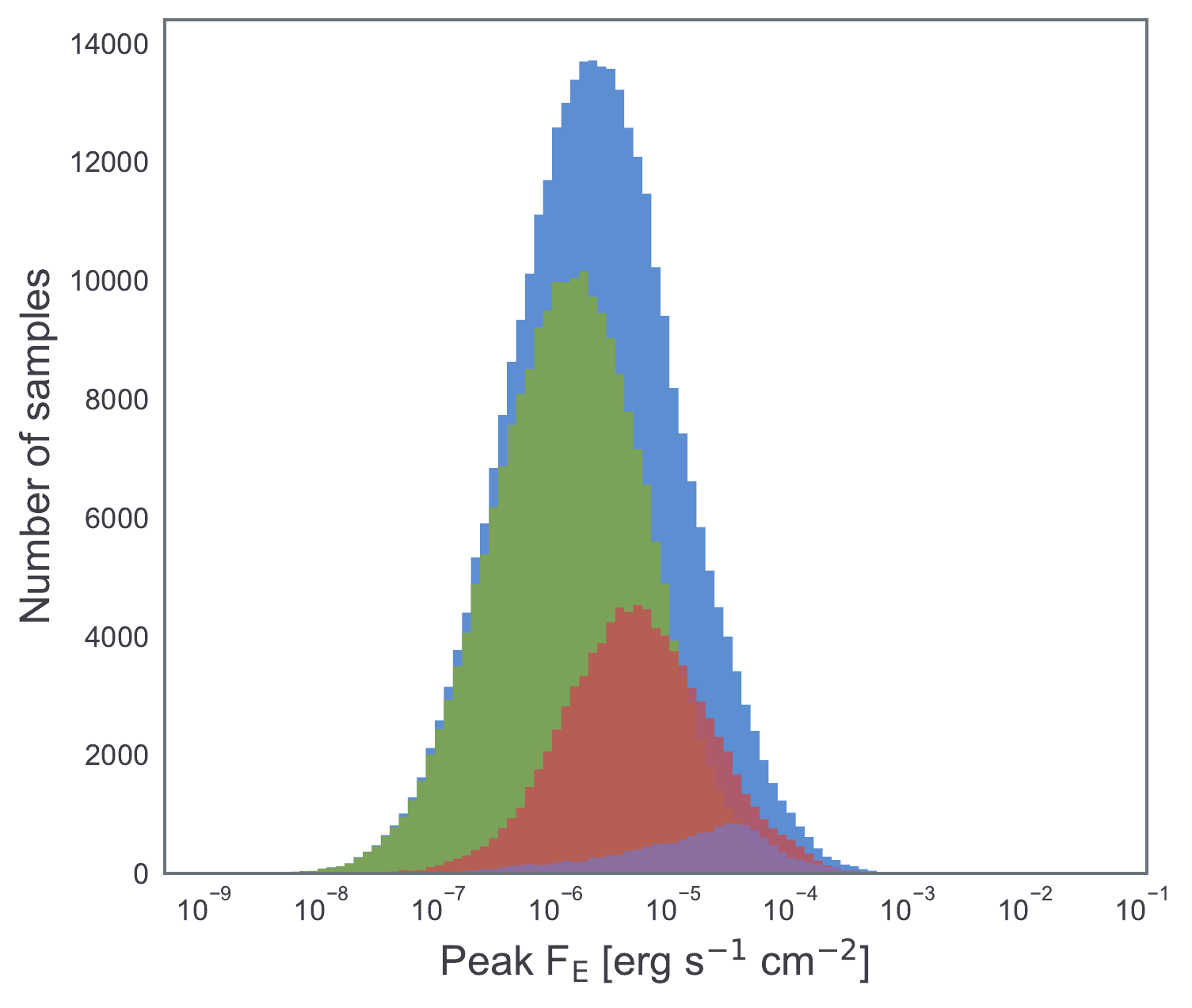}}
\caption{Combined posterior distributions for $\alpha$, E$_{\rm cut}$,
  and F$_{\rm E}$ for the peak flux spectra. Both the full sample and
  light curve structure subclasses are displayed. Note that the y-axis
  measures the number of posterior samples used.}
  
  \label{fig:comb_dist}
\end{figure*}

The marginalized parameter distributions from all GRBs can be combined
to create sample-wide distributions that fully incorporate the
individual and potentially non-Gaussian uncertainties from each
fit. In Figure \ref{fig:comb_dist}, the combined posteriors of
$\alpha$, E$_{\rm cut}$, and the F$_{\rm E}$ for the peak flux spectra
of the entire sample are displayed\footnote{While we fit for the Band
  function's high-energy spectral index ($\beta$), we do not display
  these values.}. We additionally display the distributions from the
light curve structure subclasses.

The $\alpha$ distribution of the entire sample peaks near $-2/3$ and
barely exceeds 0. The distribution is skewed towards softer values
barely exceeding $-3/2$. The behavior is generally observed in each of
the subclasses. The E$_{\rm cut}$ distribution is also skewed towards
lower values. Finally, the total peak F$_{\rm E}$ distribution is
log-normally distributed around $\sim 5 \times 10^{-5}$ erg s$^{-1}$
cm$^{-2}$. As the pulse structure moves towards complexity, the peak
of the distribution increases. This is a possible indication that lack
of pulse structure is a selection effect of the observed brightness of
the GRB. The distribution structure observed in the complex light curve subclass is
due to low sample size only.

\begin{figure*}
\centering
\figurenum{3}  \subfigure{\includegraphics[scale=.4]{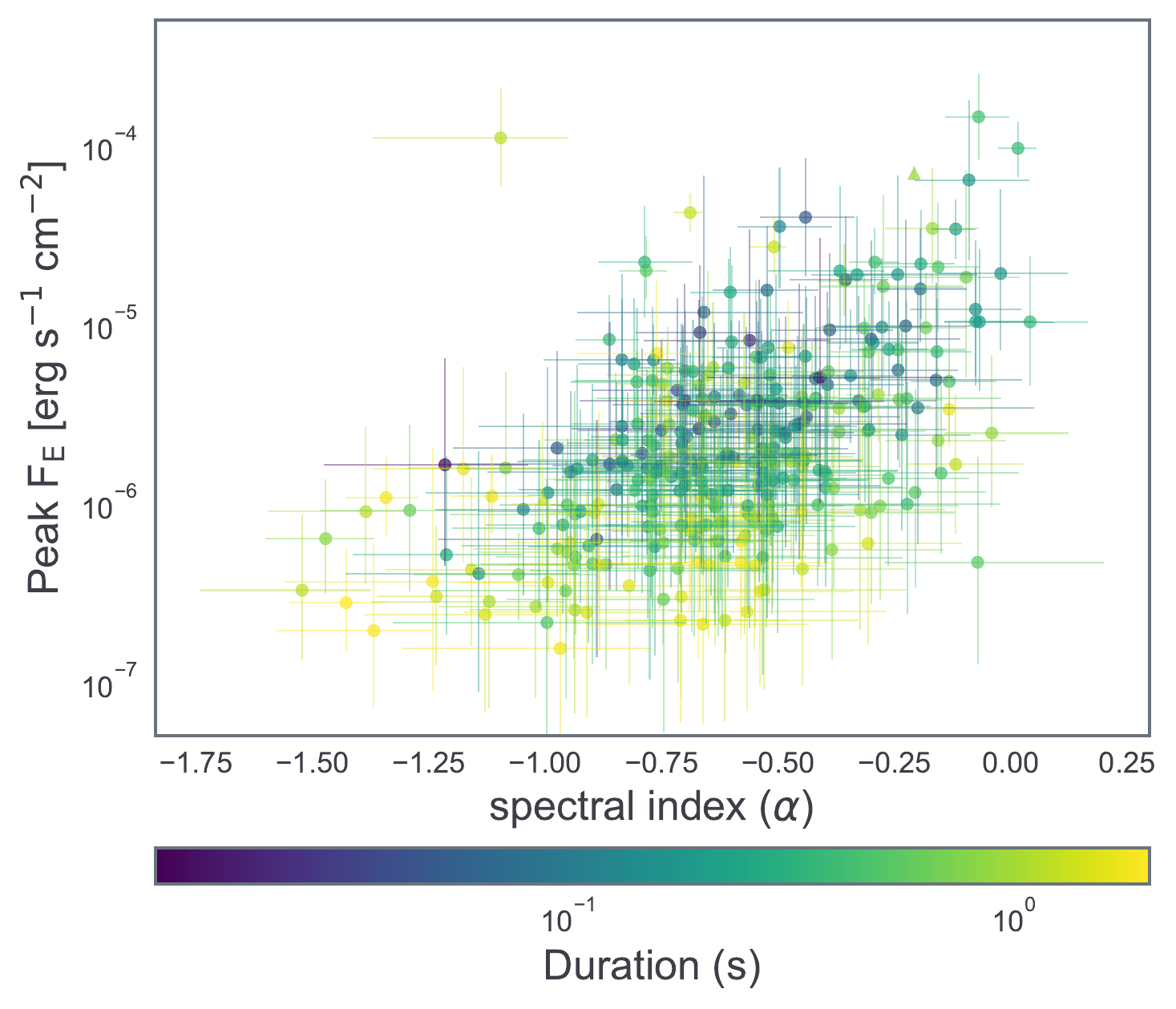}}\subfigure{\includegraphics[scale=.4]{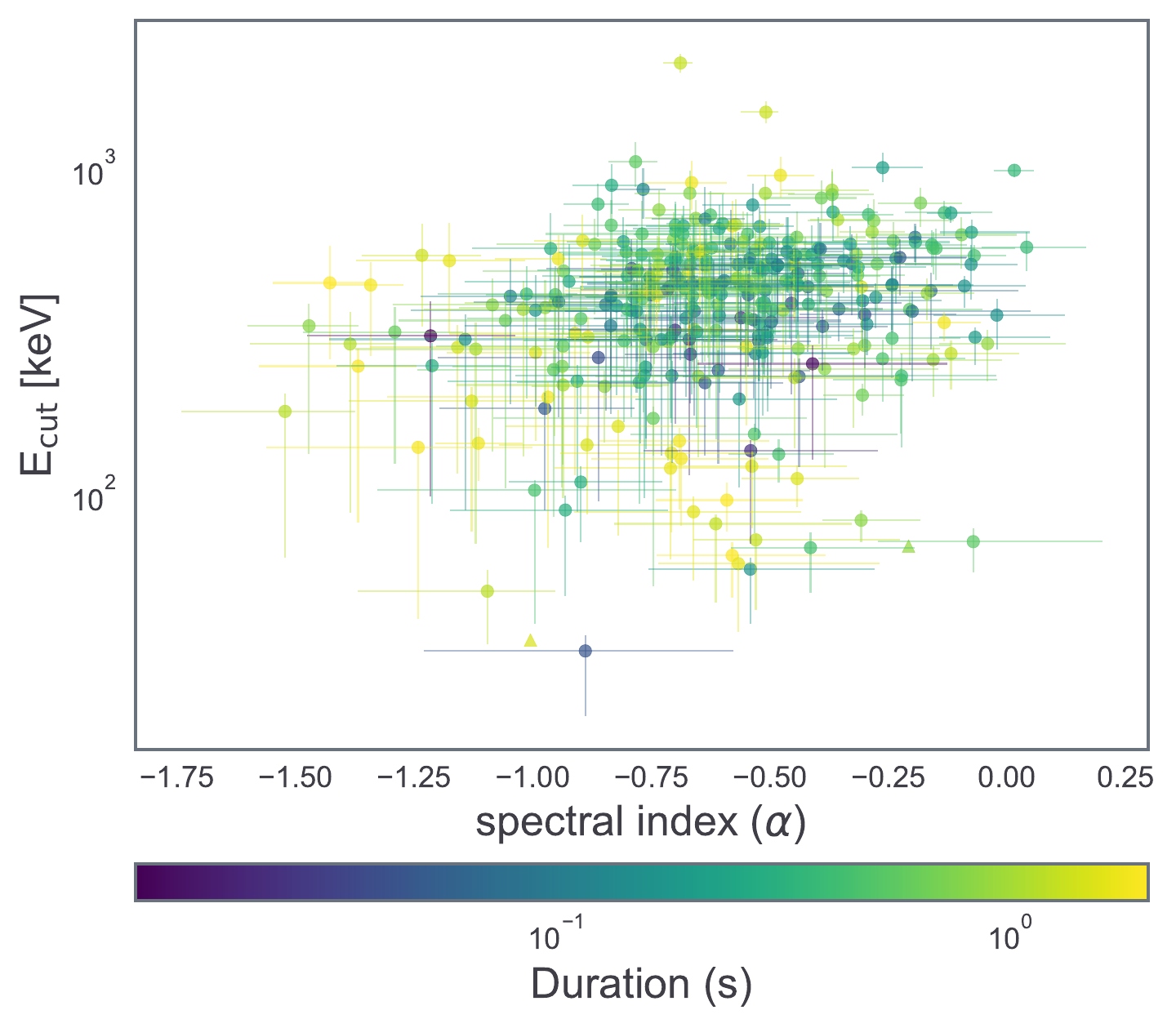}}\subfigure{\includegraphics[scale=.4]{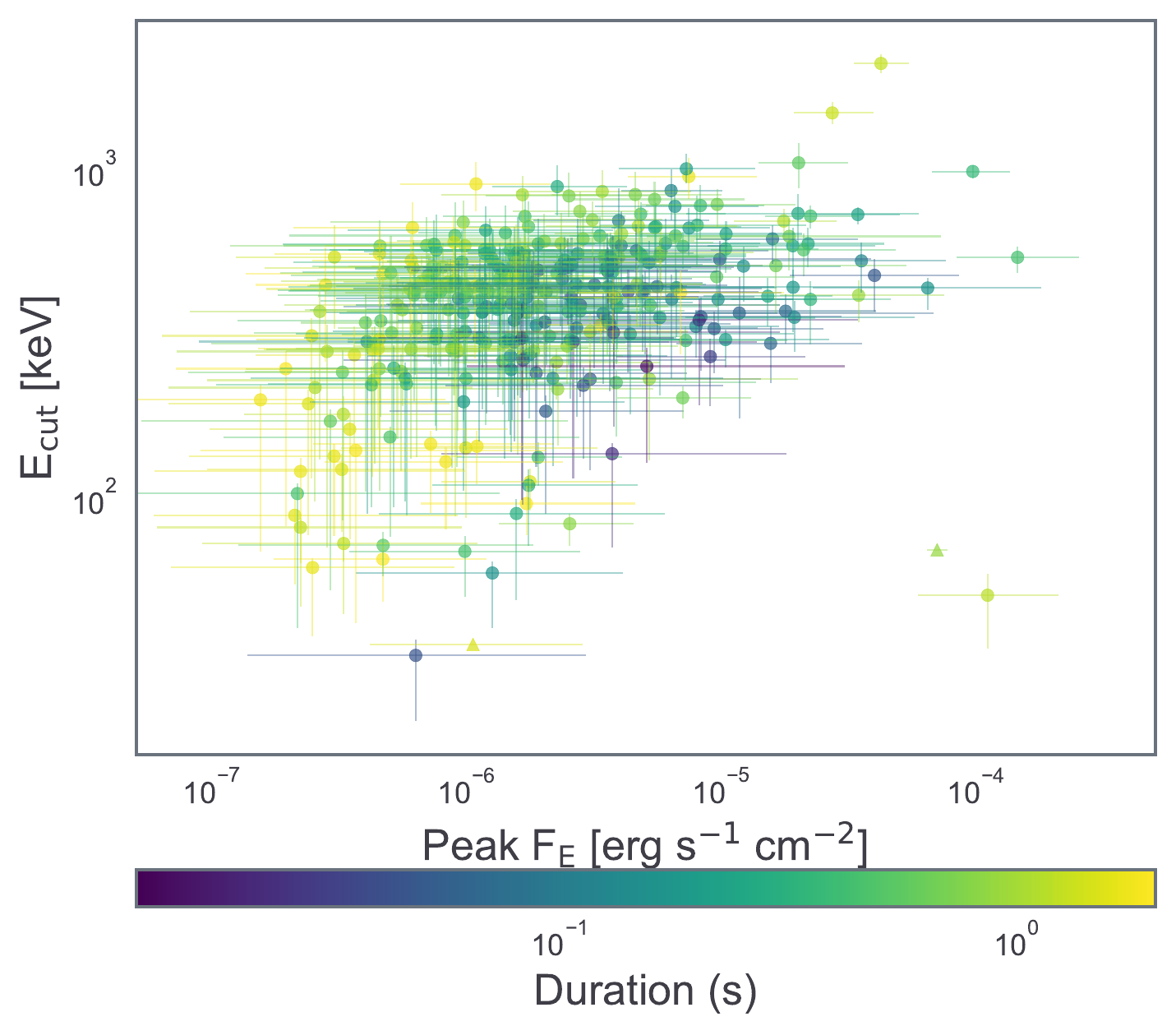}}
  \caption{From left to right the $\alpha$ - peak F$_{\rm E}$,
    $\alpha$ - E$_{\rm cut}$ and peak F$_{\rm E}$ - E$_{\rm cut}$
    distributions from the total peak flux sample. Errors are the 0.68
    credible region. The color scale indicates the duration of the
    GRB.}
  
  \label{fig:all_dist}
\end{figure*}

\begin{figure*}
\centering
 \figurenum{4} \subfigure{\includegraphics[scale=.4]{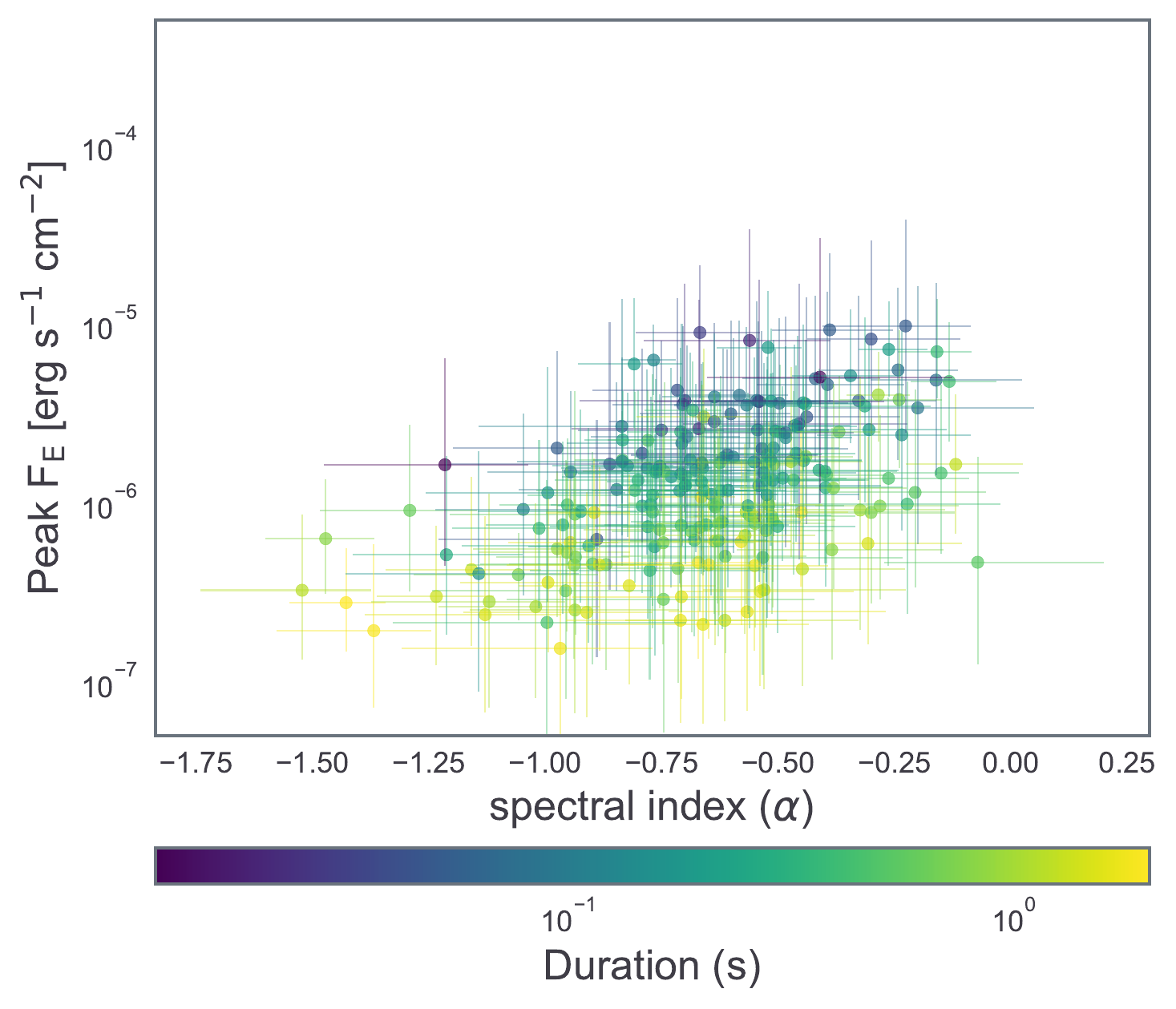}}\subfigure{\includegraphics[scale=.4]{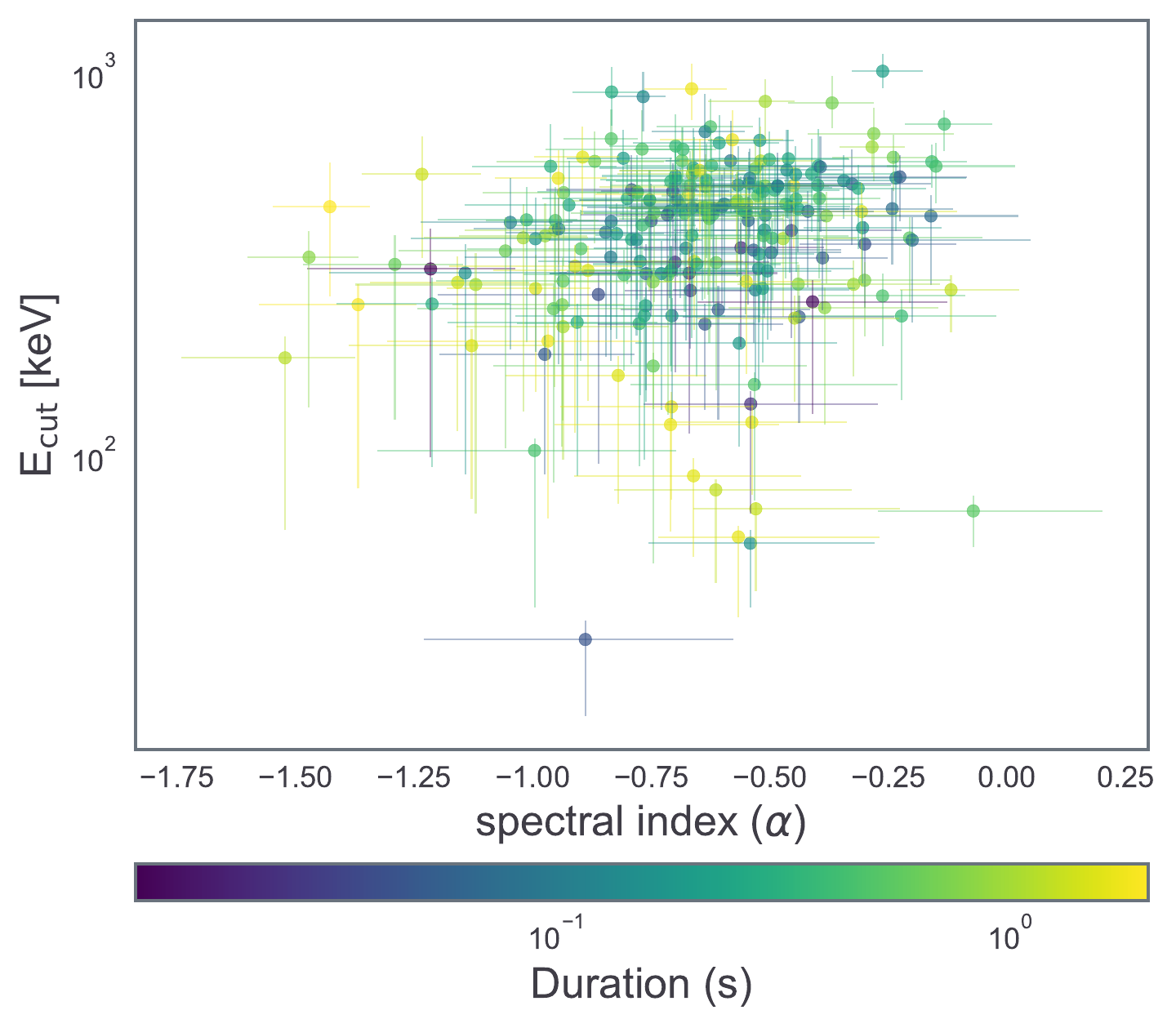}}\subfigure{\includegraphics[scale=.4]{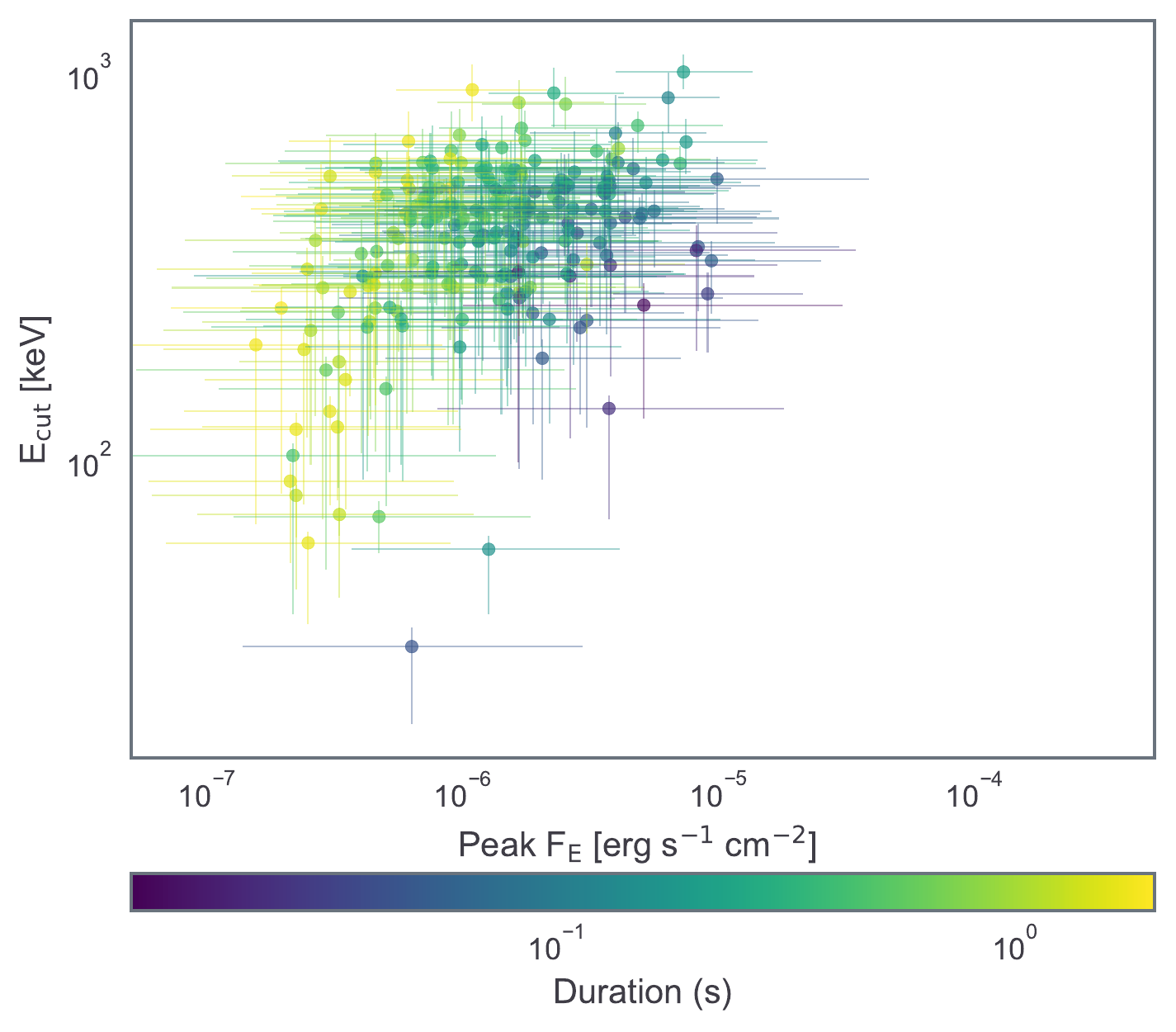}}
\subfigure{\includegraphics[scale=.4]{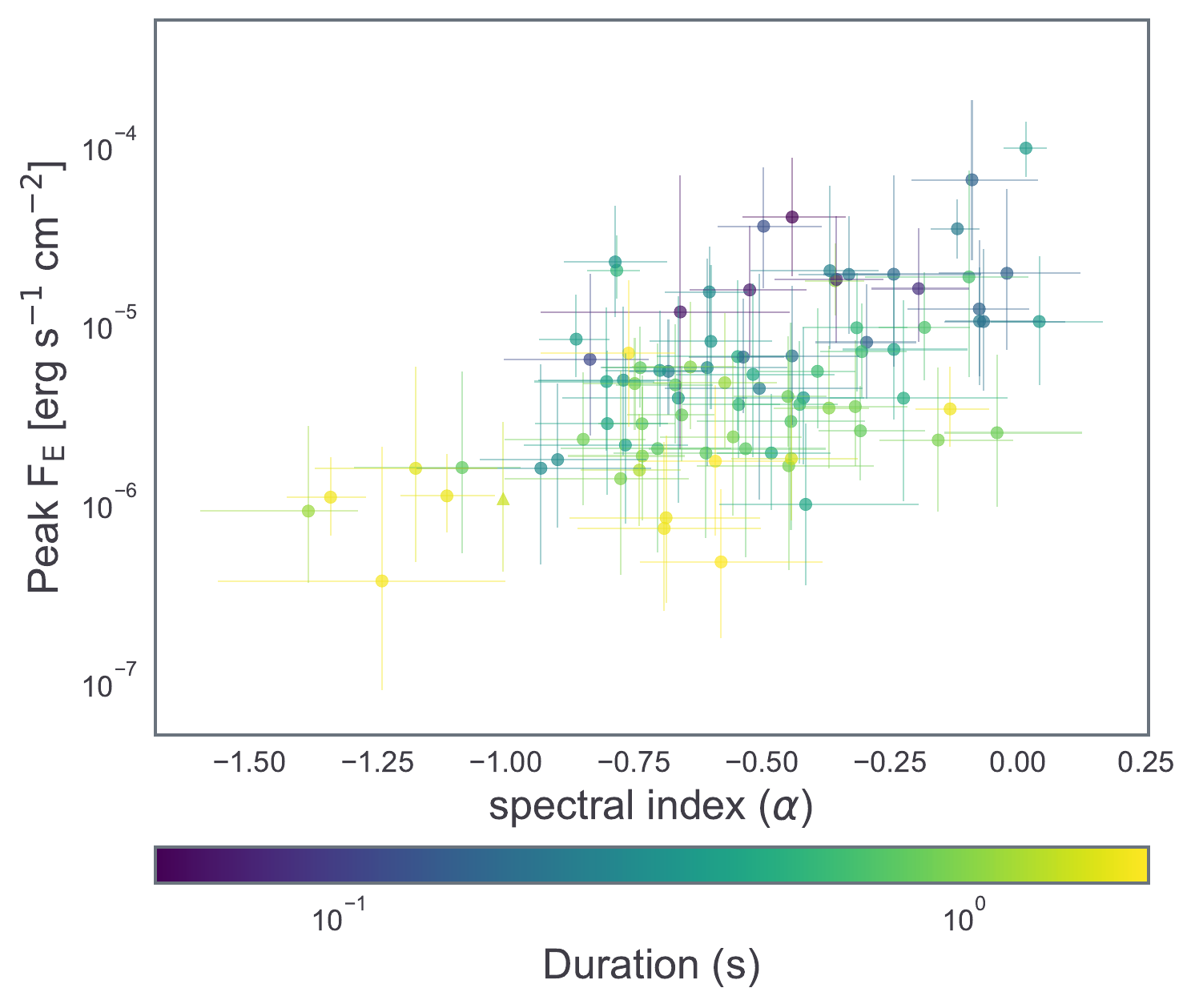}}\subfigure{\includegraphics[scale=.4]{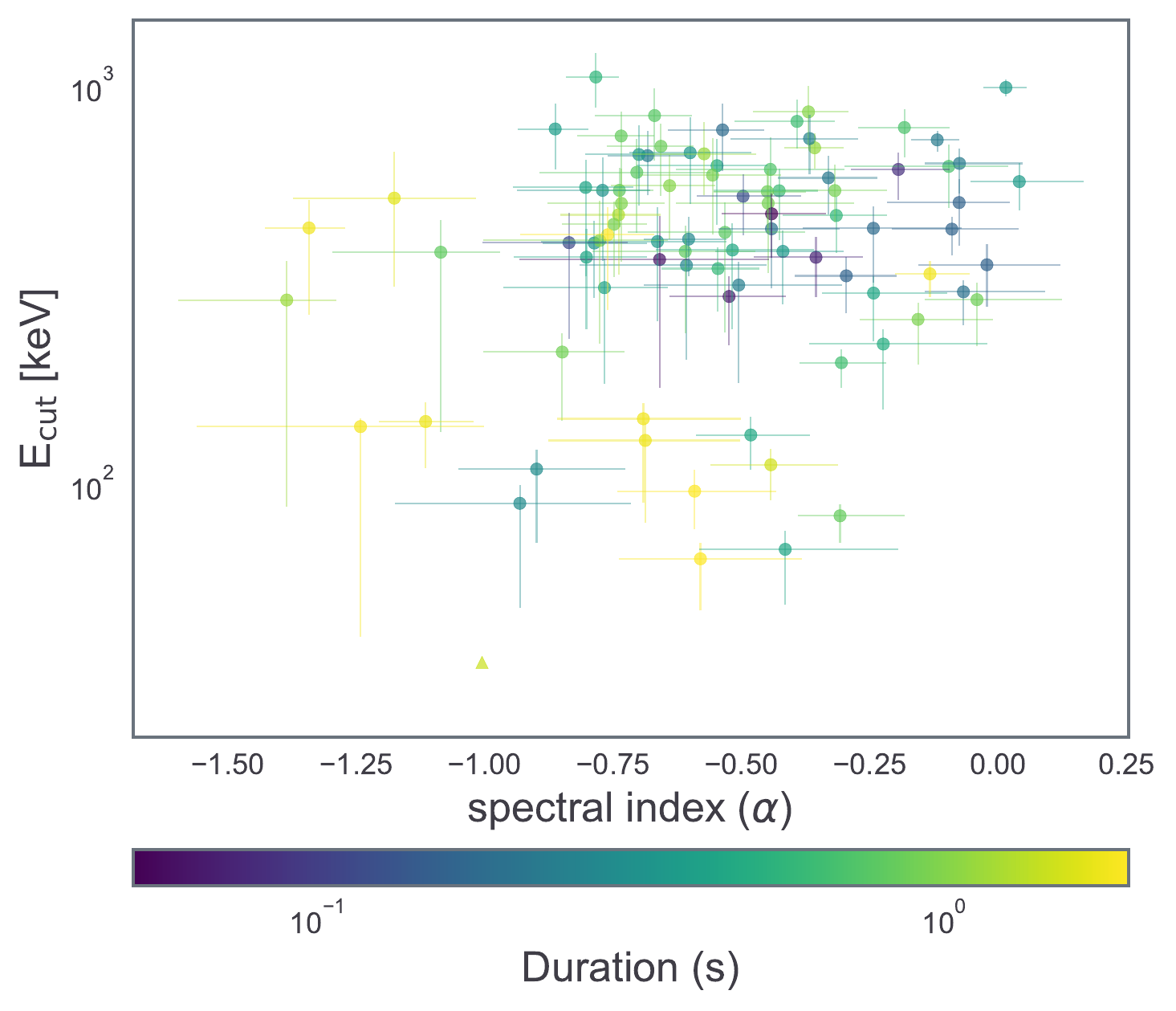}}\subfigure{\includegraphics[scale=.4]{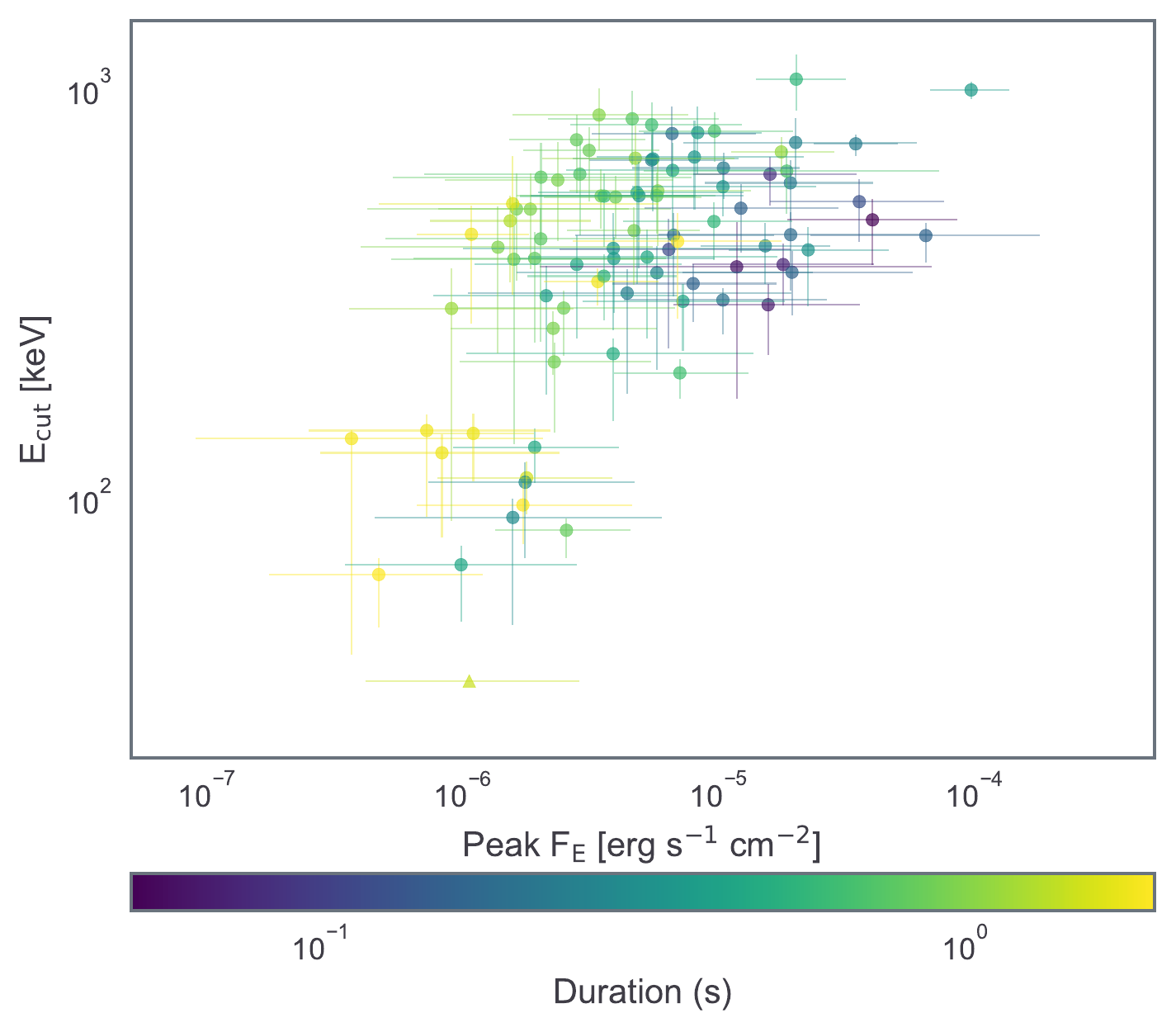}}
\subfigure{\includegraphics[scale=.4]{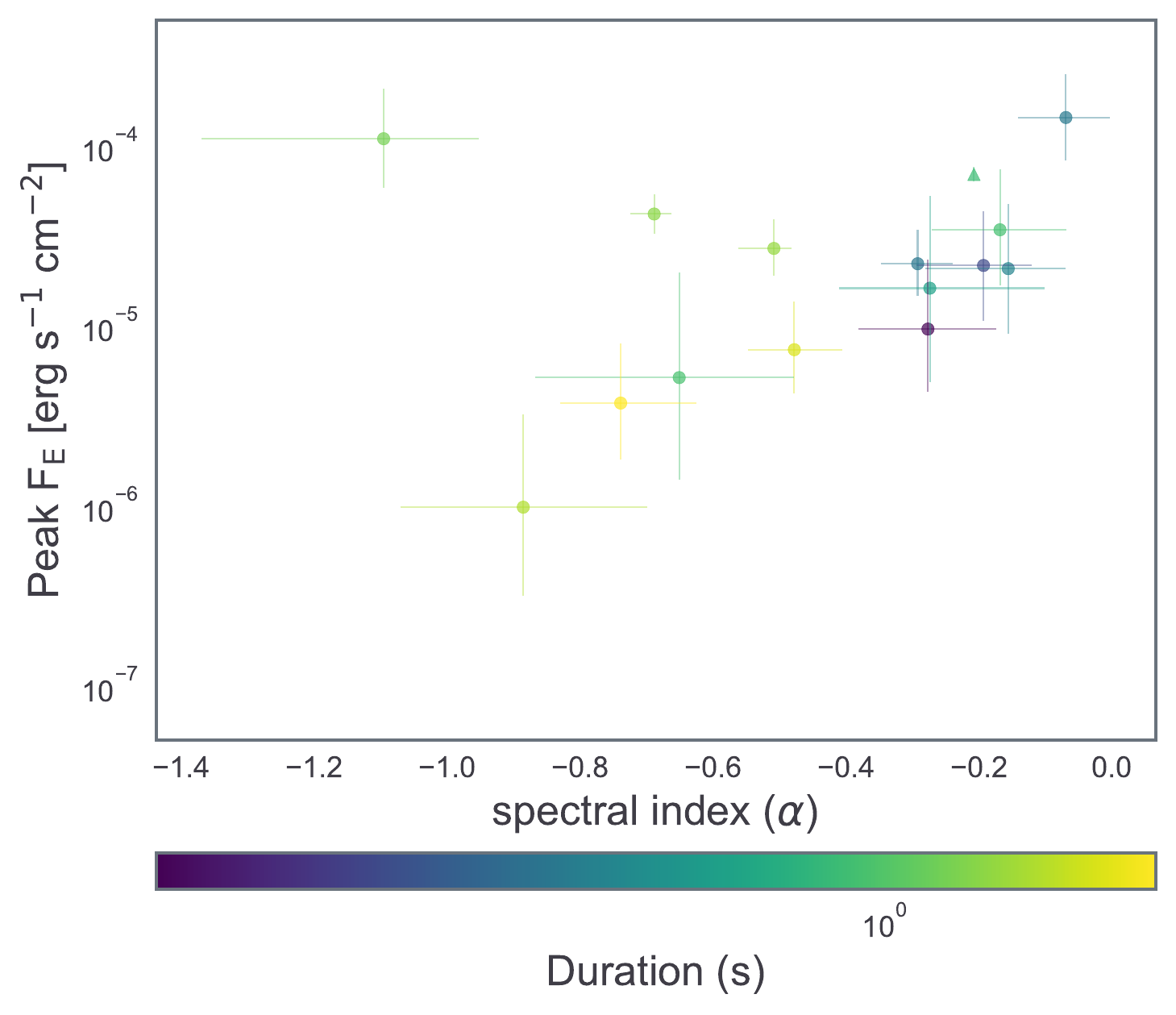}}\subfigure{\includegraphics[scale=.4]{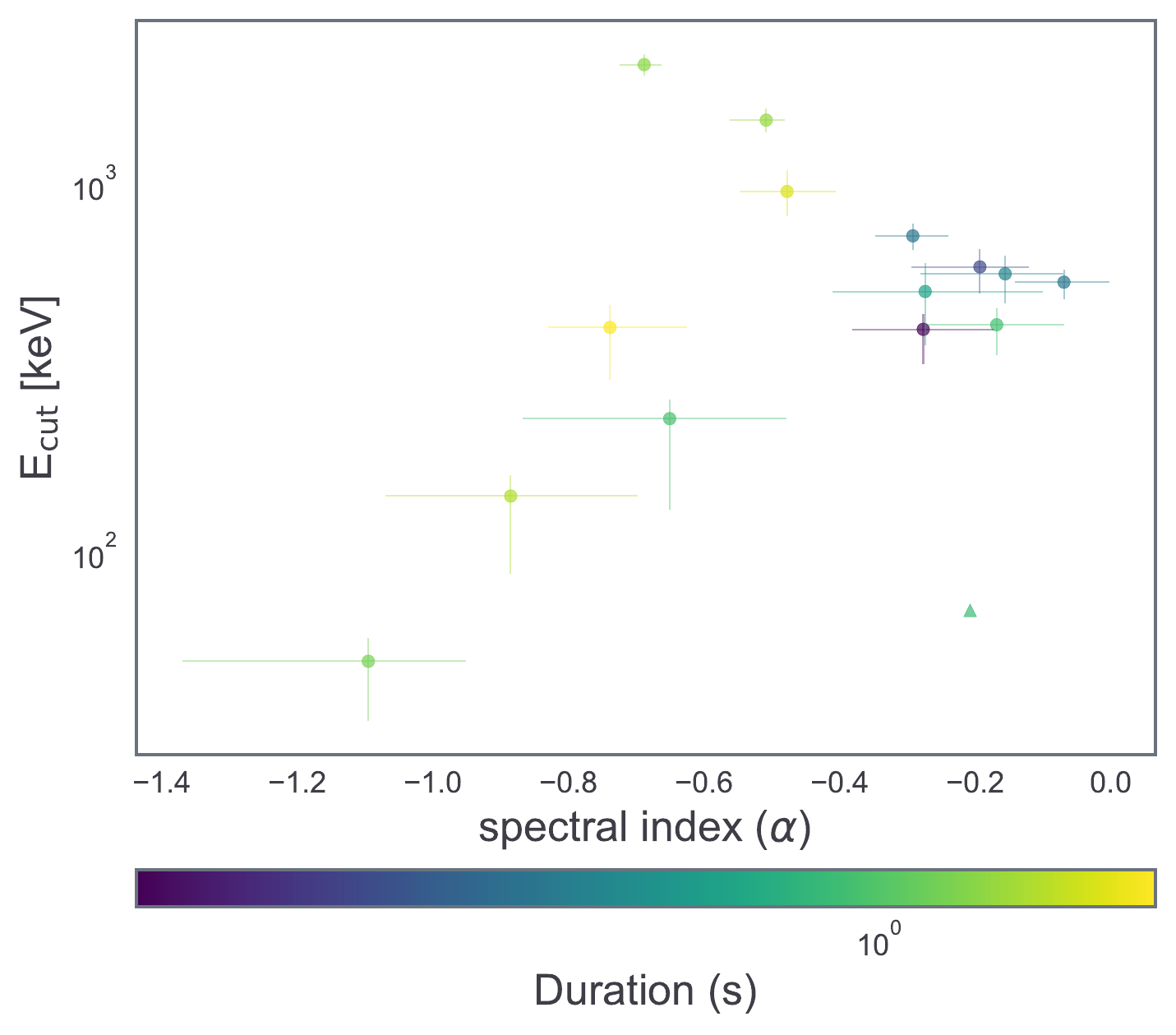}}\subfigure{\includegraphics[scale=.4]{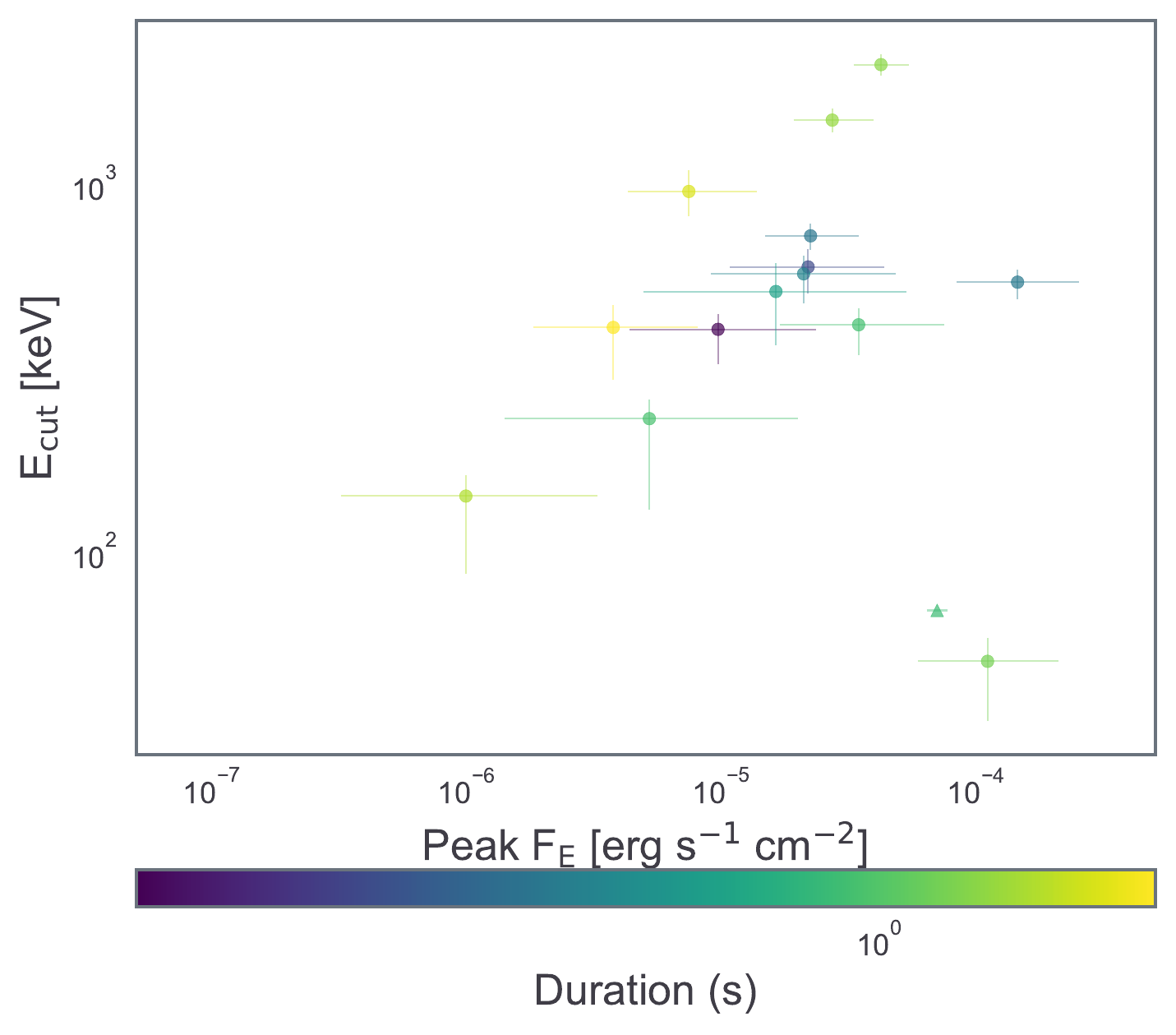}}
\caption{Same as Figure \ref{fig:all_dist} but for the simple (top),
  pulse-like (middle) and complex (bottom) light curve structure
  subclass.}
    \label{fig:sep_dist}
\end{figure*}

We also examine inter-burst parameter correlations for both the full
sample (see Figure \ref{fig:all_dist}) and all light curve structure
subclasses (see Figure
\ref{fig:sep_dist}). For
each distribution, a color scale indicating the duration of each GRB
is included. Note that we combine both GRBs best modeled by the Band
function and the CPL in these distributions and indicate with a
triangle those parameters coming from a Band function.

It is clear that for the total sample a slight correlation in
$\alpha$, peak flux and duration is observed. This becomes stronger
for the simple light curve structure subclass. However, there is no
apparent correlation between E$_{\rm cut}$ and $\alpha$. Finally,
there is a correlation between peak F$_{\rm E}$ and $E_{\rm cut}$, but
such correlations are likely attributed to functional correlations
\citep{Massaro:2007} and selections effects
\citep{Kocevski:2012}. Moreover, the discovery of GRB 170817A with a
low luminosity and low redshift should bring caution when wishing to
naively associate brightness with distance for short GRBs
\citep{Burgess:2017aa}.

\begin{figure}[h]
  \figurenum{5}
  \centering
  \includegraphics[width = 0.45 \textwidth]{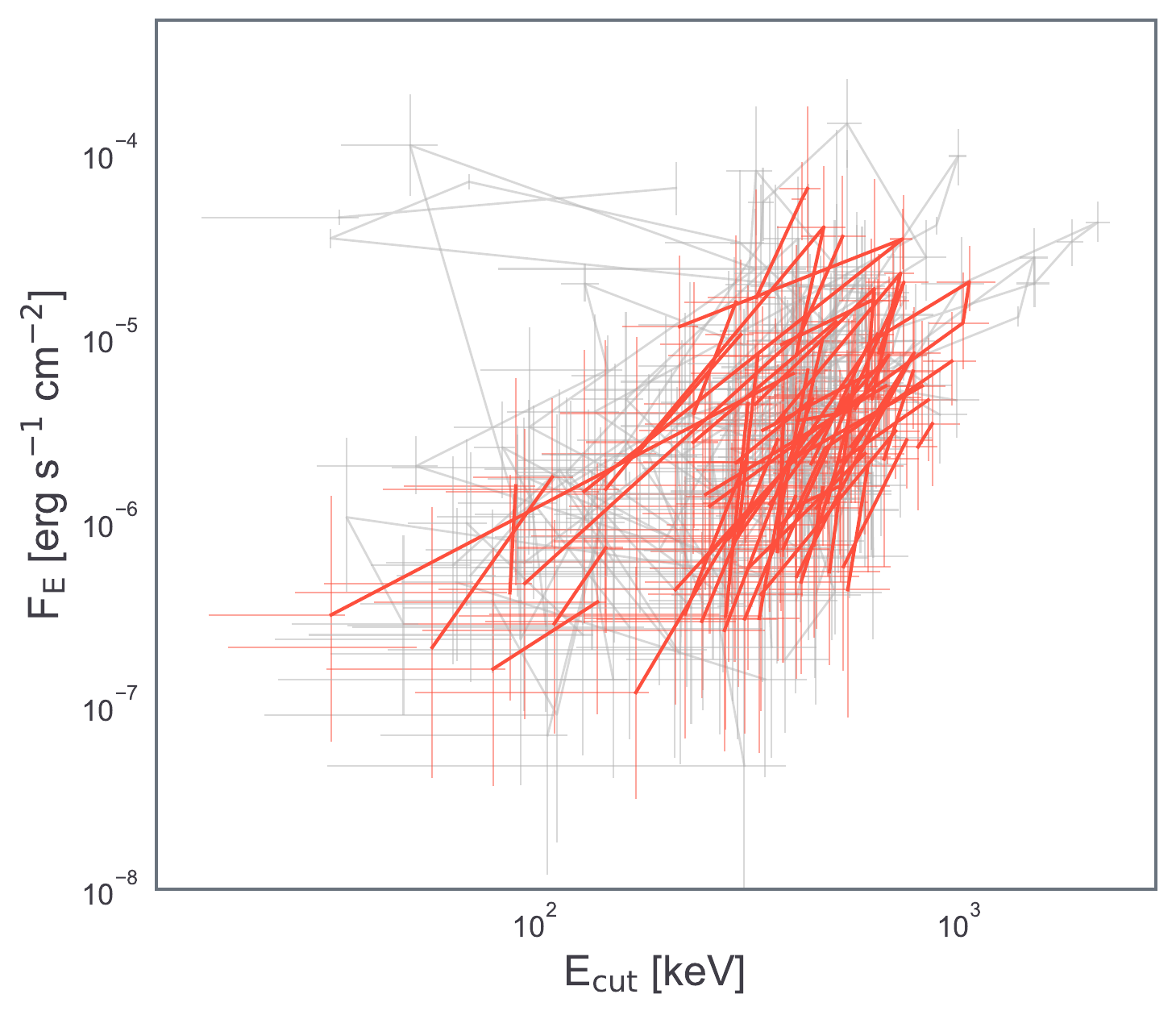}
  \caption{The E$_{\rm cut}$-F$_{\rm E}$ correlation for the
    time-resolved spectra of our sample. In red are those GRBs that
    have a strictly positive relation between the quantities.}
  \label{fig:spec_evo}
\end{figure}

Finally, we briefly examine the spectral evolution properties of our
sample. In Figure \ref{fig:spec_evo} the time-resolved F$_{\rm E}$ and
E$_{\rm cut}$ are plotted for all GRBs in our sample. Those with
strictly increasing F$_{\rm E}$ with E$_{\rm cut}$ (52 of 525) are
highlighted. The time-resolved correlation between these parameters,
first noted by \citet{Golenetskii:1983}, potentially encodes
information about the emission mechanism of the GRB. Because the
correlation is intrinsic to each GRB, it is less likely to be the
result of selection effects \citep{Ghirlanda:2010aa}. However, it is
unlikely that the correlation can be used to estimate GRB redshifts
\citep{Burgess:2015ab}.

\section{Summary}
We have presented the first fully Bayesian Fermi-GBM short GRB
catalog. In the advent of the multi-messenger era of astronomy, modern
Bayesian methodology provides a path to rigorous and sophisticated
analyses. Using the posterior distributions from our catalog allows
for non-linear error propagation of our results into further
population and emission modeling studies. Our choice of priors is
subjective, but mainly allows for us to incorporate our knowledge of
high signal-to-noise spectra into weaker spectra as well as enforcing
our belief that spectra have a cutoff in the GBM energy range. We have
checked that this does not bias our results for bright spectra where
the data become more informative than the prior. Nevertheless, as
detailed in the following section, we provide our spectral data so
that our results can be replicated and prior choices modified as seen
fit.

\subsection{Data availability}

To encourage replication and follow-up studies, we provide a variety of
data products from this study to the community. The raw spectral and
background bins are provided as PHA files readable by both 3ML and
XSPEC\footnote{https://heasarc.gsfc.nasa.gov/xanadu/xspec/}. The
spectral results are included and can be read using 3ML's {\tt
  load\_analysis\_results} function. Additionally, we include the
precomputed F$_{\rm E}$ marginal distributions. Finally, machine
readable summary tables for the time-resolved and peak flux spectral
results are released\footnote{Upon publication data will be fully
  released. Please contact the authors for access until that time.}.

\acknowledgements

The authors gratefully acknowledge the {\it Fermi} GBM team's release
of public data.  DB and FB are supported by the Deutsche
Forschungsgemeinschaft (SFB 1258).

\software{3ML, MULTINEST, pymultinest, astropy, matplotlib}

%\bibliography{bib}

\end{document}